\UseRawInputEncoding
\documentclass[twocolumn,showpacs,preprintnumbers,amsmath,amssymb,prb,aps]{revtex4-1}
\usepackage{epsfig}
\usepackage{epstopdf}
\usepackage{graphicx}
\usepackage{dcolumn}
\usepackage{bm}
\usepackage{textcomp}
\usepackage{array}
\usepackage{subcaption}
\usepackage{booktabs}

\begin{document}

\title{Exploring the suitable theoretical approach for understanding the electronic and magnetic properties of $\alpha$-Iron}

\author{Antik Sihi$^{1,}$}
\altaffiliation{sihiantik10@gmail.com}
\author{Sudhir K. Pandey$^{2,}$}
\altaffiliation{sudhir@iitmandi.ac.in}
\affiliation{$^{1}$School of Basic Sciences, Indian Institute of Technology Mandi, Kamand - 175075, India\\
$^{2}$School of Engineering, Indian Institute of Technology Mandi, Kamand - 175075, India}

\date{\today}

\begin{abstract}
We present a comparative electronic structure study using DFT and various beyond-DFT (DFT+$U$, $G_0W_0$, DFT+DMFT) methods for ferromagnetic Iron (Fe) to find better approach for describing the spectral properties of correlated magnetic system. The computed value of $U$ ($W$) is $\sim$5.4 ($\sim$0.8) eV. The calculated spectra of all methods are providing good agreement with experimental spectra (ES) for peaks' positions. But, the proper line shape is only found from DFT+DMFT with correct estimation of incoherent states, which depends on $J$ and form of local Coulomb interactions. The estimation of reduced magnetization as function of reduced temperature using DFT+DMFT shows good agreement with the experimental data. The insight of paramagnetic electronic structure of Fe is also explored. This work suggests that even for simple correlated magnetic metal, we need DFT+DMFT method to reproduce the ES with great accuracy.
       
\end{abstract}

\maketitle
\small
\section{Introduction} 

  Nowadays, the progress in modern technology demands newly predicted materials with exotic phenomena. In light of this, strongly correlated electron systems (SCES) are very famous for showing diverse phenomena like heavy fermionic behavior \cite{coleman2001,gegenwart2008}, spin and cluster-glasses \cite{binder1986}, non-Fermi liquid property \cite{stewart2001,antikprb2020}, Kondo insulator \cite{fisk}, topological properties \cite{lu2013,pdutta}, etc. The full many-body Hamiltonian equations of these systems are not exactly solvable using any present theoretical procedure. Thus in present days generally, there are two ways to get the approximate values of eigenfunctions and eigenvalues of these systems \cite{nilsson2018}. In first approach, generally few bands are considered for calculating some particular part of the full Hamiltonian, like Anderson impurity model \cite{anderson}, Hubbard model \cite{hubbard} and so on. This type of theoretical approach, which is known as model Hamiltonian based method, is very famous due to less computational complexity and has good success for providing better explanation of many experimental observations of localized electrons ($i.e$ $d$ or $f$ orbitals' electrons). But, different model needs some material specific parameters for carrying the calculation. For example, in case of the multiorbital Hubbard model for $d$/$f$ electrons, at least three different parameters, which are hopping integral ($t$), on-site Coulomb interaction ($U$) and exchange integral ($J$), need to be provided. Thus, when we have experimental data of any material for comparison then these parameters can be varied to get the better matching with the experiment. But, this is not possible for the prediction of new material, where experimental data are absent. In such situation the second approach comes to provide us parameter free full electrons method, which are called as first-principles based method. Density functional theory (DFT) is a famous first-principles based method \cite{jones1989}. In this technique, the full Hamiltonian is mainly approximated by introducing different exchange correlation (e.g. LDA \cite{lda}, GGA\cite{gga}, SCAN\cite{scan} etc.). This method has great success to describe the electronic structure of the nearly free-electron-states ($i.e.$ $t$ $\gg$ $U$), where $s$ or $p$ orbitals' electrons are occupied in outer most orbitals of the material. But, many times this method shows improper metallic ground state for many known insulating materials \cite{anisimov1997,himmetoglu2013,gopal2017}. In that case, Green's function based full electrons $GW$ ($G$ = one particle Green's function and $W$ = fully screened Coulomb interaction) method is good to provide proper ground state for weakly ($t$ $\gg$ $U$) or moderately ($t$ $\approx $ $U$) correlated materials \cite{hedin,martin,kutepov}. This technique has the ability to tackle the screening effect within Hartree-Fock approximation by using many-body perturbation theory. The bandgaps of many semiconductors, where the top most occupied orbitals are filled by $s$ or $p$ orbitals' electrons, are nicely estimated by this method \cite{oshikiri,markvan,sakuma,jiang2012}. However, it is known that these methods do not properly explain many exotic phenomena of the SCES ($t$ $\ll$ $U$).

   In order to describe both the nearly free electrons and correlated localized electrons with good accuracy, the combination of model Hamiltonian and first-principles based methods (combined method) is preferable. It is noted that the SCES are commonly formed with these type of combined electrons. Although, this combination within the two methods creates problem due to double-counting (DC) of electronic energy, which is now smartly tackled by developing different DC scheme. But, for calculation we need to provide the values of interaction parameters $U$ and $J$, which will create hopeless situation for predicting any new material. Thus, to overcome this problem, there are two widely used first-principles based methods for calculating the material specific values of $U$ and $J$. These two methods are known as constrained density functional theory (cDFT) \cite{anisimov1991,paromita1,paromita2} and constrained random-phase approximation (cRPA) \cite{aryastiawan2004,vaugier2012}. Many times, it is found that cDFT calculates unnecessary larger value of $U$ due to not considering the screening effect, which is typically not observed in cRPA \cite{karlsson2006,miyake2008}. Most commonly used combined method for electronic structure calculation of any material is DFT+$U$ \cite{anisimov1997}, where the calculated values of $W$ and $J$ from cRPA calculation are preferred to use. This method provides quite good explanation of the experimental spectra (ES) as evident from our previous work \cite{antik}. However, it is beyond the scope for providing information about finite temperature incoherent excitations of material. Our previous work on paramagnetic (PM) Vanadium also showed that proper estimation of incoherent states were important for proper line shape of the calculated spectra (CS) as compared to ES \cite{antik}. This behaviour can be possible to get by using DFT+dynamical mean field theory (DMFT) method, which is one of the advanced theoretical method and also classified into one of the combined method \cite{georges,kotliar,yee}. In this method, the dynamics of the correlation effect is tackled. It is found that the estimation of magnetic transition temperature ($T_C$) depends on the form of Coulomb interaction \cite{belozerov}. Han et al. showed in their work that different form of Coulomb interactions within DFT+DMFT method predicts different values of $T_C$, but all overestimate the experimental value\cite{qhan}. In addition to this, the value of $T_C$ also depends on the choice of impurity solver \cite{belozerov2013}. However, in present days, Continuous-time quantum Monte Carlo (CT-QMC) is known as the better choice for the quantum impurity solver in DFT+DMFT calculation. In case of magnetic system, it is known that $J$ is an important interaction parameter. Therefore, proper estimation of material specific $J$ is very important because it is seen that the value of $T_C$ varies with changing the value of $J$ \cite{belozerov}. Hence, keeping in mind all those above discussion, it is really necessary to perform a comparative study for finding the better theoretical methods to investigate the magnetic correlated electron systems. Therefore, Iron (Fe) is the perfect candidate for this case study.

\begin{figure}
  \begin{center}
    \includegraphics[width=0.70\linewidth, height=5.0cm]{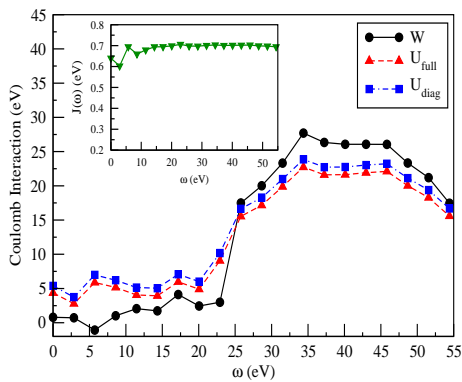} 
    \caption{\small Coulomb interaction as function of $\omega$. }
    \label{fig:}
  \end{center}
\end{figure}

  This elemental metal is a very famous itinerant-magnet from the ancient time due to its' various usefulness in daily life. The magnetic DFT calculation provides quantitative explanation of low-temperature ferromagnetic (FM) properties of Fe \cite{singh1991,singh1994}. But, due to presence of correlation in 3$d$ electrons, proper estimation of $U$ and $J$ are also very important for this metal. In general for 3$d$ materials, the value of $J$ is similar to its' atomic value of 0.9 eV \cite{belozerov,zaanen}. But, in case of Fe, the value of $J$ is typically found in between 0.65 to 0.75 eV, which is evident from the recent studies by using cRPA method \cite{miyake2008,sasioglu2011}. A wide range of values of $U$ (1.0 eV to 6.0 eV) and $J$ (0.9 eV to 1.0 eV) are used for studying different physical properties of this material as found from literature \cite{belozerov,uandj1,uandj2,uandj3,uandj4,uandj5}. However, from Auger spectra, the effective $U$ between two 3$d$ holes are measured to be 1 to 2.6 eV \cite{yin1977,treglia1981}. Different choices of values of $U$ and $J$ provide good descriptions of different experimental observations for this elemental metal using DFT+DMFT method \cite{belozerov,uandj1,uandj2,uandj3,chadov2008}. Thus, a huge range of values for $U$ and $J$ creates a hopeless situation for describing individual physical properties of Fe with varying different combination of $U$'s and $J$'s values. Therefore, one of the best way to verify the validation of this interaction parameters may be done by comparing the calculated density of states (DOS) with ES. To the best of our knowledge, there are few works for Fe, where this comparison is carried out only for valence band (VB) with X-ray-photoemission spectroscopy (XPS) data \cite{fink2009,minar2011}. But, the comparison of conduction band (CB) with bremsstrahlung isochromat spectroscopy (BIS) data still needs to be studied along with VB for establishing the better predictive power of any electronic structure method.

  In our present study, the importance of $J$ for studying the spectral properties of this transition metal, comparative study of different $ab$ $initio$ methods for increasing predictive strength of individual method and understanding the electronic structure of PM phase are focused. The values of $U$, $J$ and $W$ for this transition metal are estimated from the cRPA calculation and found to be in good agreement with the previous studies. The $\omega$ dependent Coulomb interactions show the importance of correlation effect and the presence of plasmon excitation for this metal. Here, the comparative study with different electronic structure methods ($i.e.$ DFT, DFT+U, one-shot $GW$ ($G_0W_0$) and DFT+DMFT) suggests that the proper estimation of electronic states and line shape of ES can only be done by DFT+DMFT method with the proper choice of Coulomb interaction and the value of $J$. In order to get the proper explanation of ES by DFT+DMFT method, the major role of $J$ instead of $U$ is found for this transition metal. For exploring the insight of coherent and incoherent state of Fe, momentum resolved spectral function is calculated at 300 K. The calculated value of magnetization of this magnetic metal at 200 K is found to be $\sim$2.28 ($\sim$2.25) $\mu_B$/Fe for Full (FullS) Coulomb interaction, which is nicely matched with the experimental data. Furthermore, the electronic structure of PM $\alpha$-Fe is explored in details by using DFT+DMFT method.
 
\begin{figure}
  \begin{center}
    \includegraphics[width=0.98\linewidth, height=7.8cm]{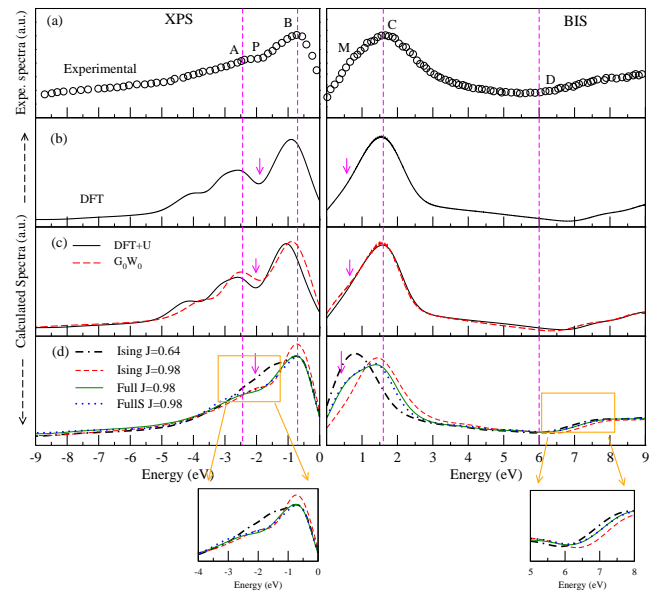} 
    \caption{\small (a) Experimental valence band using XPS \cite{ley} and conduction band using BIS \cite{speier}, calculated spectra using (b) DFT, (c) DFT+$U$ \& $G_0W_0$ and (d) DFT+DMFT (at 300 K) in ferromagnetic phase. Zero energy represents the Fermi level.}
    \label{fig:}
  \end{center}
\end{figure}

\section{Computational details}

 To study the electronic structure of Fe, the spin-polarised calculation is performed using the full-potential linearized-augmented plane-wave method. The lattice parameter's value of 2.8665 \AA \, is used with space group of $I$m-3m \cite{chichagov}. The convergence criteria for total energy is fixed at 10$^{-4}$ Ry/cell for 10 $\times$ 10 $\times$ 10 k-mesh. Here, DFT calculation with LDA is carried out by WIEN2k code \cite{wien2k}. cRPA method is chosen for calculating the different Coulomb interactions. These calculations are performed by using GAP2 code with the Wannier basis function \cite{jiang2016,jiang2013}. This code is also utilized for $G_0W_0$ calculation. Elk code is used for DFT+$U$ due to its' simple and general implementation \cite{elk,bultmark}. The calculations of DFT+DMFT are done by using eDMFT\cite{yee} code for different temperatures, where WIEN2k\cite{wien2k} code is utilized for the self-consistent DFT calculation through out the DMFT iterations. The CT-QMC is chosen to solve this impurity problem with the \textquoteleft exactd\textquoteright \, DC method \cite{khaule,hauleprl}. The analytical continuation is done to obtain the spectral function in real axis by using maximum entropy method \cite{jarrell}.

\section{Results and Discussion} 

 The electronic bands of Fe metal near to the Fermi energy ($E_F$) are mainly formed by 3$d$-orbitals, which are selected for eliminating the 3$d$-3$d$ electronic transitions to find the values of $U$ and $J$ using cRPA. The calculated values of intra bare Coulomb interaction, intra fully screened Coulomb interaction and intra Coulomb interaction ($U_{intra}$) are 22.99 (22.13) eV, 0.61 (0.92) eV and 5.47 (5.32) eV for $e_g$ ($t_{2g}$) orbitals, respectively. The value of inter Coulomb interaction ($U_{inter}$) for $e_g - e_g$ ($t_{2g} - t_{2g}$) orbitals is estimated to be 3.89 (3.99) eV. Moreover, the computed values of $U_{inter}$ for $d_{z^2} - d_{xy}$, $d_{z^2} - d_{xz}$, $d_{z^2} - d_{yz}$, $d_{x^2-y^2} - d_{xy}$, $d_{x^2-y^2} - d_{xz}$ and $d_{x^2-y^2} - d_{yz}$ are 3.85 eV, 4.38 eV, 4.38 eV, 4.56 eV, 4.03 eV, 4.03 eV, respectively. The value of full Coulomb interaction ($U_{full}$) (diagonal Coulomb interaction ($U_{diag}$)) is found to be $\sim$4.4 ($\sim$5.4) eV, which is calculated by taking the average value of all (diagonal) matrix elements of the Coulomb interaction. The diagonal unscreened (bare) Coulomb interaction, full bare Coulomb interaction, on-site bare exchange interaction, $J$ and $W$ are computed to be 22.47 eV, 21.25 eV, 0.77 eV, 0.64 eV and 0.8 eV, respectively. Moreover, the values of $J$ and $W$ are nicely agreed with the previously calculated values by cRPA method \cite{miyake2008,sasioglu2011,karlsson2006,miyake2006}.

\begin{figure}
  \begin{center}
    \includegraphics[width=0.85\linewidth, height=6.0cm]{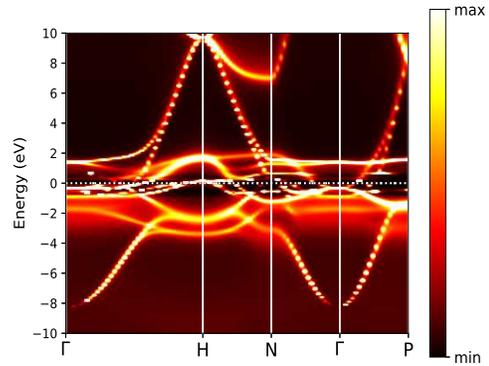} 
    \caption{\small Momentum-resolved spectral function of Iron metal in ferromagnetic phase at 300 K.}
    \label{fig:}
  \end{center}
\end{figure} 

  To study the importance of orbital screening effect, the $\omega$ dependent $U_{full}$, $U_{diag}$ and $W$ are plotted in Fig. 1. The figure shows slightly higher values of $U_{diag}$ than $U_{full}$ for the observed $\omega$ range. In higher $\omega$ region, the calculated values of these interactions are seen to be closer to the value of bare Coulomb interaction, which denotes the less importance of electronic screening in higher $\omega$. However, below 26.0 eV, these values of all Coulomb interactions are rapidly decreased. This scenario indicates the importance of screening effect in lower $\omega$, where the values of $U_{full}$ and $U_{diag}$ are found to be much higher than the value of $W$. Hence, this calculation suggests the significance of excluding the 3$d$-3$d$ electronic transitions to obtain the $U$ for this metal. The minimum value of $W$ is found to be $\sim$-1.1 eV, which is obtained at $\sim$5.7 eV. This negative value of $W$ may be due to the plasmon excitation, where the intra 3$d$ band electrons' transitions are responsible for getting this behaviour. This value is found within the typical range ($i.e.$ 5 to 15 eV) of plasmon excitation of any metal. In addition to this, the $\omega$ dependent $J$ is also shown in the inset of Fig. 1, where the values of $J(\omega)$ are not significantly changing with $\omega$. The values of $U_{diag}$, $W$ and $J$ at $\omega$=0.0 eV are used for the further calculations.

  Here, the value of $U_{diag}$ at $\omega$=0.0 eV represents the value of Hubbard-$U$ (=5.4 eV), which shows good matching with the previous works \cite{karlsson2006,nakamura2006}. The ES obtained from XPS (BIS) data provide the information of the VB (CB). Therefore, if the CS from any electronic structure calculation are nicely matched with these ES, then the predicted physical quantities by the corresponding theoretical method will be reliable. In this work, the convoluted total DOS (TDOS), which represents the CS, is obtained with the similar procedure as mentioned in our earlier work \cite{antik}. The different theoretical methods are used to obtain the CS, which are shown in Fig. 2 along with the ES ($i.e$ XPS\cite{ley} and BIS\cite{speier}). Here, Fig. 2(a) shows the ES, where two peaks in VB and one peak in CB are observed. The peak at $\sim$-2.44 ($\sim$-0.7) eV is marked by A (B). Moreover in between peaks A \& B, a plateau like feature is indicated by P. The peak height ratio between peaks A to B is measured to be $\sim$0.7. In CB, the peak is found at $\sim$1.6 with a mound like feature within 0.5 eV to 1.1 eV. The mound like feature and the peak are marked by M and C, respectively. To discuss the monotonically increasing nature in ES observed after 6.0 eV, a dashed line D is drawn at 6.0 eV.

  At first one of the most commonly used electronic structure method, DFT is chosen for comparing CS with the ES. The CS using DFT are plotted in Fig. 2(b). In VB of this CS, the positions of the first two peaks are found to be almost similar with the peaks A \& B. But, a dip (marked by an arrow) instead of plateau like feature is observed and one hump at $\sim$-4.0 eV is also seen from this CS. However, this hump is not found in the ES. The peak's position in CB of CS is nicely matched with the peak C. The monotonically increasing behaviour is seen after $\sim$7.0 eV. The mound like features in CB is not observed from this CS. Therefore, it is evident from this comparison that DFT is not able to provide proper explanation of ES except the major peaks' positions. The earlier discussion on different $\omega$ dependent Coulomb interactions suggests to pay special attention for tackling the electronic correlations effect of Fe 3$d$ electrons. In case of DFT, the electronic correlations for localized electrons are not sufficiently considered, which may be the reason to get improper line shape of CS. Thus, it is expected that DFT+$U$ method, where DFT is combined with Hubbard model Hamiltonian, may provide the better estimation of ES.

\begin{figure}
  \begin{center}
    \includegraphics[width=0.70\linewidth, height=5.0cm]{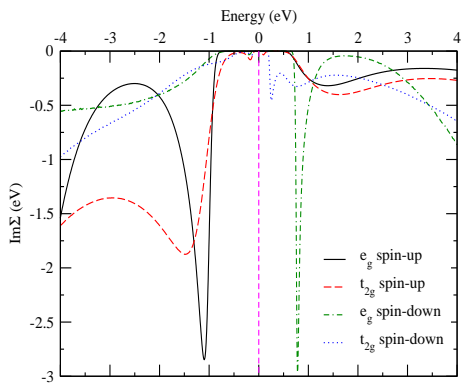} 
    \caption{\small Imaginary part of self-energy at 300 K in ferromagnetic phase.}
    \label{fig:}
  \end{center}
\end{figure}

  The proper choice of material specific $U$ is very important for DFT+$U$ calculation. From our previous work\cite{antik}, it is concluded that if the value of $W$ is used as the $U$-parameter, then DFT+$U$ gives better explanation of ES. Therefore, in this work, the $U$ value of 0.8 eV, which is equal to the value of $W$, is used for DFT+$U$. The CS obtained from this calculation are shown in Fig. 2(c), where the line shape, peaks' positions and peaks' height are almost same as DFT except the peak B and shallowness of dip. The peak corresponding to peak B is shifted by $\sim$0.35 eV towards the lower energy and the shallowness of the dip is slightly improved than DFT. But, still more improvements are needed for the better matching of line shape. However, it is expected that the theoretical method, which considered the electronic interaction as function of \textbf{k} and $\omega$, may give the good matching in between CS and ES. But, DFT and DFT+$U$ are not categorized into this type of method, whereas modern $GW$ based technique may tackle these problems. Therefore, the CS using $G_0W_0$ method are computed and plotted in Fig. 2(c). All the peaks' positions are fairly good matched with the ES. The figure shows that shallowness of the dip is reduced than DFT and DFT+$U$, but the other features are found to be similar with those methods. Thus, this method also fails to provide the proper explanation of ES. However, it is known that the amount of electronic correlation considered in $G_0W_0$ method is not sufficient for SCES. Hence, it will be interesting to see whether DFT+DMFT has ability to describe the proper explanation of ES or not for this simple correlated metal.

 \begin{figure}
  \begin{center}
    \includegraphics[width=0.75\linewidth, height=7.5cm]{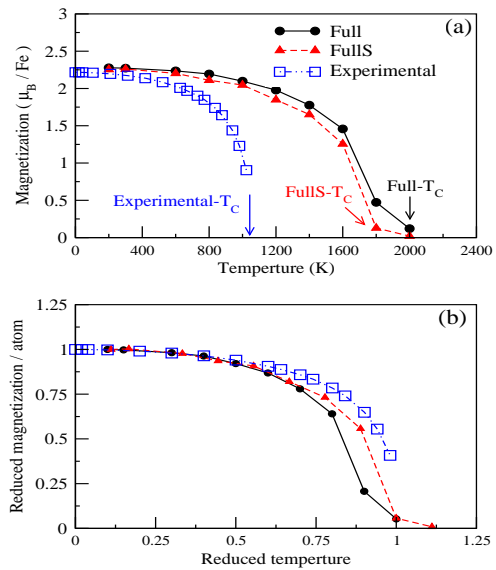} 
    \caption{\small (a) The temperature dependence magnetization of $\alpha$-Iron using different Coulomb interactions along with experimental data \cite{crangle}, (b) reduced magnetization as function of reduced temperature along with experimental data \cite{crangle}.}
    \label{fig:}
  \end{center}
\end{figure}

  The value of $U_{diag}$ is chosen as $U$-parameter because electrons screening between 3$d$-3$d$ electrons are already considered in DFT+DMFT method. Therefore, the $U$ value of 5.4 eV is used for this calculation with the estimated value of $J$ (= 0.64 eV) by using cRPA method. The CS using these values with considering density-density (Ising) type local Coulomb interaction are shown in Fig. 2(d). The figure represents that the CS using these interaction parameters are providing worst explanation of the ES, except the positions of peak B and the monotonically increasing nature after $\sim$6.0 eV. The $T_C$ value for this metal strongly dependents on the value of $J$, but not very much depends on $U$ \cite{belozerov}. Therefore, the Yukawa screening, which is a different procedure to calculate the value of $J$, is used and the value found to be $\sim$0.98 eV using $U$=5.4 eV. This computed value of $J$ is very close to previously calculated values using cDFT \cite{anisimov1991,belozerov}. The CS using these values of $U$ and $J$ with Ising interaction are plotted in Fig. 2(d). The figure shows that the peaks' positions, plateau and monotonically increasing nature are in good agreement with the ES, except the peak related to peak A is slightly shifted towards the higher energy. However, still more improvements are needed for explaining the ES because the mound like behaviour in CB and the proper line shape of plateau are missing. Now, whether the improvement of line shape in CS depends on the form of Coulomb interaction or not, we performed DFT+DMFT calculations with $U$=5.4 eV and $J$=0.98 eV by two different type of Coulomb interactions. The first one is fully rotationally invariant Coulomb interaction (Full) with $SU(N)$ symmetry as similar to Kanamori type and the second one is fully rotationally invariant Coulomb interaction of Slater type (FullS). The CS using the Full, which is considered to be more improved form of Coulomb interaction than Ising, are illustrated in Fig. 2(d). In overall, the line shape, peaks' positions, plateau like nature and the monotonic increasing in CS are showing fairly good agreement with the ES. Particularly, the mound like behaviour in CB is clearly seen from this CS, which may be due to spectral weight transfer from coherent to incoherent states. Afterwards, the similar calculation with FullS is performed for computing the CS, which are plotted in Fig. 2(d). All the features are found to be similar with the CS of Full, except the line shape at the plateau region are improved further. Therefore, at this point, it is noted that FullS is adequately explaining the ES. Moreover, in the inset of Fig. 2(d), all CS obtained by using DFT+DMFT are plotted for more clear visualization of the improved line shape around the plateau and the more clear comparison of monotonically increasing behaviour of CS. Thus, from this comparative study, the importance of the proper estimation of states is understood for the better matching of ES. In this scenario, DFT+DMFT with fixed $U$=5.4 eV and $J$=0.98 eV provides proper estimation of states by using FullS. At this point, it is interesting to note that the value of $U$ calculated by cRPA method is good to explain the ES but not the value of $J$. In case of $J$, the predicted value by cDFT is quite good. Therefore, this result suggests that the improvement of cRPA is needed for calculating $J$ value of magnetic system.  

 \begin{figure}[]
   \begin{center}
   \begin{subfigure}{0.85\linewidth}
   \includegraphics[width=0.80\linewidth, height=4.5cm]{fig6a.eps}
   \caption{}
   \label{fig:} 
\end{subfigure}
\begin{subfigure}{0.85\linewidth}
   \includegraphics[width=0.95\linewidth, height=6.0cm]{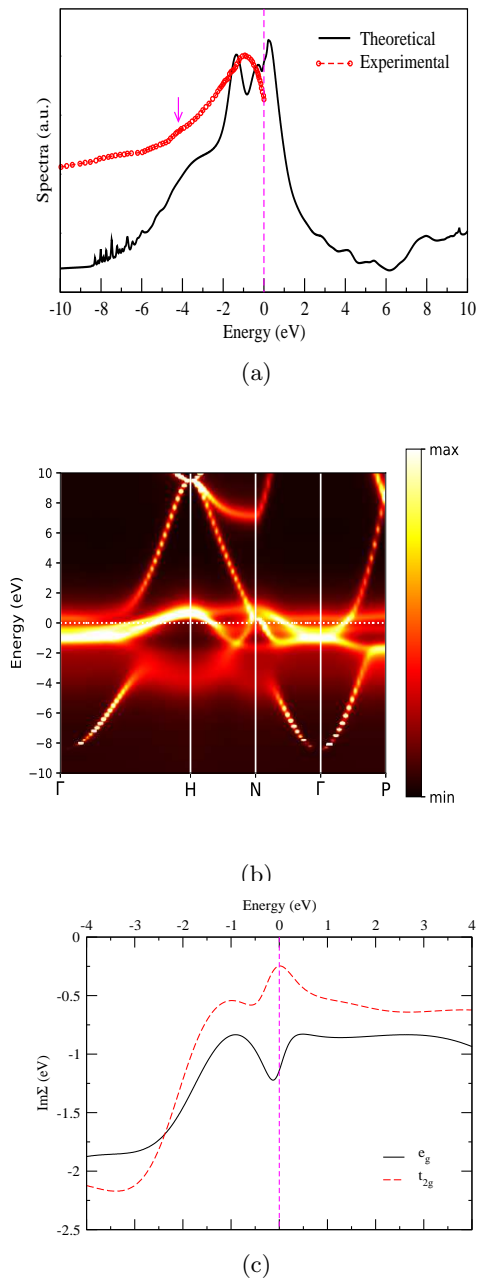}
   \caption{}
   \label{fig:}
\end{subfigure}
\begin{subfigure}{0.85\linewidth}
   \includegraphics[width=0.80\linewidth, height=4.5cm]{fig6c.eps}
   \caption{}
   \label{fig:}
\end{subfigure}
\caption{\small (a)Theoretical spectra along with experimental spectra \cite{mcalister1972}, (b) momentum-resolved spectral function and (c) imaginary part of self-energy at 1800 K in paramagnetic phase using DFT+DMFT method. Zero energy represents the Fermi level.}
   \end{center}
\end{figure}

  Now, to understand whether the incoherent or coherent states are responsible for providing the nicely matched line shape of CS as compare to ES, the momentum-resolved spectral function in FM needs to be explored. Therefore, it is calculated using DFT+DMFT with FullS interaction at 300 K along the $\Gamma$-H-N-$\Gamma$-P direction as shown in Fig. 3. In this plot, the minimum (min) to maximum (max) colour range represents the increment of coherency for the quasiparticle states. Therefore, here the reddish (yellowish) colour denotes the incoherent (coherent) states. It is observed from the figure that the presence of incoherent states are clearly seen at $\sim$-2.0 eV energy region along this observed high symmetric k-directions. This may be the reason to get proper line shape of plateau from this method. Similarly in CB, the incoherent states are found between $\sim$0.3 eV to $\sim$1.5 eV along this k-directions. This feature may be responsible for getting the mound like feature in CS as similar to ES. The existence of incoherent states around these mentioned energy region can also be possible to verify from the imaginary part of self-energy ($Im\, \Sigma (\omega)$).

  The orbital dependent $Im\, \Sigma (\omega)$ as a function of energy is plotted in Fig. 4 by using FullS for -4.0 to 4.0 eV. The spin-up (spin-down) of the $e_g$ and $t_{2g}$ orbitals show very small value of $Im\, \Sigma (\omega)$ from $\sim$-0.7 ($\sim$-0.13) eV to $\sim$0.5 ($\sim$0.6) eV and $\sim$0.05 ($\sim$0.08) eV to $\sim$0.02 ($\sim$0.16) eV, respectively, as evident from the figure. It suggests the presence of coherence state of spin-up (spin-down) $e_g$ and $t_{2g}$ orbitals in this energy window. At $\sim$-2.0 eV, where the plateau nature is observed in ES, all the spin channel with their corresponding orbitals are showing sufficiently large values of $Im\, \Sigma (\omega)$. This large amount of $Im\, \Sigma (\omega)$ represents the existence of high number of incoherent states for these orbitals, which are responsible for providing good line shape in CS at the plateau region. In addition to this, $\sim$0.2 eV to $\sim$1.0 eV, spin-down of $e_g$ and $t_{2g}$ orbitals are major contributor of the large value of $Im\, \Sigma (\omega)$. Thus in CS, the population of incoherent state for spin-down $e_g$ and $t_{2g}$ orbitals are giving the mound like shape in CB near to $E_F$. Moreover, the value of $m^*$ is given by, 
  
\begin{equation} 
\frac{m^*}{m_{DFT}} = 1 - \frac{dRe\, \Sigma (\omega)}{d\omega}\mid_{\omega=0}
\end{equation}
where, $Re\, \Sigma (\omega)$ is the real part of self-energy. The values of $\frac{m^*}{m_{DFT}}$ for spin-up (spin-down) of the $e_g$ and $t_{2g}$ orbitals are found to be $\sim$1.70 ($\sim$1.53) eV and $\sim$1.96 ($\sim$1.81) eV, respectively.

  Experimentally, it is known that FM to PM transition for Fe occurs at temperature 1043 K \cite{crangle}. Here, the magnetization ($M$) of this magnetic metal is calculated for different temperatures by DFT+DMFT with Full and FullS, which are shown in Fig. 5(a) along with experimental \cite{crangle} data. The experimental value of saturation magnetization at zero temperature is $\sim$2.217 $\mu_B$/Fe. Here, the value of $M$ at 200 K from DFT+DMFT calculation is found to be $\sim$2.28 ($\sim$2.25) $\mu_B$/Fe for Full (FullS), where sufficient saturation in temperature dependent magnetization curve is seen. Thus, it shows a fairly good agreement with the experimental data. But above 300 K, it is observed that the deviation between experimental and calculated values of $M$ is rapidly increasing with rise in temperature. The Full interaction are showing higher values than the FullS for the observed temperature range. The $T_C$ is estimated to be $\sim$1800 K ($\sim$2000 K) from the DFT+DMFT using FullS (Full), where the experimental value is 1043 K \cite{crangle}. Thus, DFT+DMFT overestimates the $T_C$ for both these Coulomb interactions, but whether the decrement nature of calculated $M$ with temperature follows the similar behaviour as found from experiment or not, the reduced magnetization obtained from calculations (Full and FullS) are plotted as function of reduced temperature along with experimental data in Fig. 5(b). Here, the reduced magnetization is obtained when the magnetization of corresponding temperature is divided by the saturation magnetization of corresponding method. In similar way, the reduced temperature is obtained when the temperature is divided by the corresponding estimated $T_C$. The figure shows nicely matching till $\sim$0.6 reduced temperature between two theoretical curves and the experimental curve. Afterwards, the deviation between these curves are observed. The variation between the curves of Full and experiment is slightly higher than the curve of FullS. Moreover, the curvature of the theoretical curve obtained by FullS is in fairly good agreement with the experimental curve. Hence within DFT+DMFT method, the FullS provides reasonably good explanation not only the ES but also the behaviour of the magnetic transition of this transition metal. However, to get better matching for the value of $T_C$, further improvement in Coulomb interaction is needed.
  
\begin{figure}
  \begin{center}
    \includegraphics[width=0.75\linewidth, height=5.0cm]{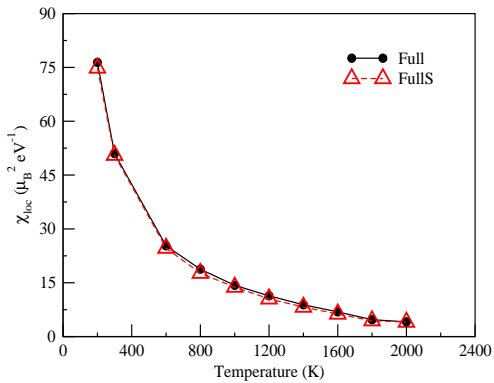} 
    \caption{\small Temperature dependent local spin susceptibility ($\chi_{loc}$) at $\omega$=0 by different form of local Coulomb interactions.}
    \label{fig:}
  \end{center}
\end{figure}

  Now, to explore the electronic structure of Fe in PM phase, the non-magnetic DFT+DMFT calculation is carried out at 1800 K with FullS form of Coulomb interaction. The theoretical spectra (TS) obtained from this calculation are shown in Fig. 6(a) along with ES of VB. From ES, the main peak is observed at $\sim$-1.0 eV with a hump at $\sim$-4.2 eV. In case of VB for TS, the major peak's and one hump's positions (marked by an arrow) are found to be $\sim$-1.3 eV and $\sim$-3.4 eV, respectively. However, this hump is not observed in the spectra of FM phase. At this point, it is interesting to note that the positions of this peak and hump are in quite good agreement with experimental data \cite{mcalister1972}. Now in CB, the another major peak is seen at $\sim$0.22 eV. The increasing trend in TS for CB is observed after 6.0 eV. Alongside with this, to get more detailed information about the nature of electronic excited states at this temperature, momentum-resolved spectral function along high symmetric k-direction is plotted in Fig. 6(b). The figure illustrates that the states near to $E_F$ are mainly contributed by the coherent states except along H-N direction and near to P-point. However, the presence of incoherent states around the $E_F$ is also observed along the high symmetric k-direction, which is quite large in number as compared to FM phase. Moreover, it is obvious to get such large amount of incoherent behaviour in high-temperature region. Below $\sim$-1.5 eV, the incoherent states is seen from the figure. But above 2.0 eV, the similar nature of coherent and incoherent states are observed both in FM and PM phases. To achieve more insight of the incoherent features of the states at 1800 K, $Im\, \Sigma (\omega)$ for $e_g$ and $t_{2g}$ orbitals are plotted in Fig. 6(c) for -4.0 to 4.0 eV. From -2.0 eV to 2.0 eV, the major contribution of $Im\, \Sigma (\omega)$ is coming from $e_g$ orbitals as evident from the figure. In this energy window, the values of $Im\, \Sigma (\omega)$ for $e_g$ ($t_{2g}$) orbitals are estimated to be from $\sim$-0.8 ($\sim$-0.25) eV to $\sim$-1.5 ($\sim$-1.25) eV. Therefore, such a large values of $Im\, \Sigma (\omega)$, which are found from these two orbitals, suggest the presence of high amount incoherent states of both $e_g$ and $t_{2g}$ orbitals within this energy window. Moreover, the calculated values of $\frac{m^*}{m_{DFT}}$ for $e_g$ and $t_{2g}$ orbitals are $\sim$1.72 eV and $\sim$2.19 eV, respectively, at 1800 K in PM phase. At last, the temperature dependent local spin susceptibility ($\chi_{loc}$) at $\omega$=0.0 is calculated for Full and FullS to study the local spin-spin correlations, which are shown in Fig. 7. Hausoel et al. show in their work that the calculated values of $\chi_{loc}$ at $\omega$=0 by Ising and Full type interactions have sufficient difference in their observed temperature range. But, from the figure, Full and FullS are showing almost similar values of $\chi_{loc}$ at $\omega$=0 for all observed temperatures. The values of $\chi_{loc}$ at $\omega$=0 for 200 K found to be $\sim$76.4 ($\sim$74.8) $\mu_B^2\, eV^{-1}$ in case of Full (FullS). Finally in FM phase, the average free electrons in 3$d$ orbitals is found to be $\sim$6.26 at 200 K, where the major probable electronic configurations are coming from $d^5$, $d^6$ and $d^7$. It represents the mixed electronic configurations of this transition metal. Similar behaviour in electronic configurations is also seen from PM phase of $\alpha$-Fe.

\section{Conclusions} 

 In this study, a detailed comparative electronic structure calculations using both of first-principles ($i.e.$ DFT and $GW$) and combined method ($i.e.$ DFT+$U$ and DFT+DMFT) are carried out for Fe. The calculated value of $U$ ($W$) using cRPA is $\sim$5.4 ($\sim$0.8) eV. In order to describe the spectral properties of this transition metal, the importance of correlation effect is found to be evident from the $\omega$ dependent Coulomb interactions. Different $ab$ $initio$ methods provide nicely matched peaks' positions compare to experimental spectra (ES), except DFT+DMFT with calculated value of $J$ by cRPA. Although, the proper description of line shape as similar to ES is only seen from DFT+DMFT calculation with FullS form of local Coulomb interaction using $J$=0.98 eV, which is calculated by considering Yukawa screening with fixed $U$=5.4 eV. The correct estimation of incoherent states is found to be important for getting the proper line shape. The magnetization values estimated by DFT+DMFT with Full and FullS are nicely matched with experimental data below 300 K. But, the estimated value of $T_C$ is quite higher than experimental data, whereas the reduce temperature dependent reduced magnetization curve using FullS nicely follows the experimental curves. Importance of many-body interaction effect to study the paramagnetic electronic structure of Fe is also found by comparing ES with theoretical spectra. Thus, all these result suggest the importance of $J$ and essential needs of combined first-principles with many-body impurity model based method ($e.g.$ DFT+DMFT) for describing the ES of this simple correlated metal. Alongside with this, the failure of cRPA method for calculating material specific $J$ is also found for this simple magnetic material.  

\section{References}


\end{document}